\documentclass[aps,prd,preprintnumbers,groupedaddress,nofootinbib]{revtex4}

\pdfoutput=1

\usepackage{graphicx}
\usepackage{latexsym}
\usepackage{amsfonts}
\usepackage{amssymb}
\usepackage{amsmath}
\usepackage{slashed}
\usepackage{array}
\usepackage{feynmp}
\usepackage[hypertex]{hyperref}
\usepackage{url}

\def\lsim{\mathrel{\raise.3ex\hbox{$<$\kern-.75em\lower1ex\hbox{$\sim$}}}}
\def\gsim{\mathrel{\raise.3ex\hbox{$>$\kern-.75em\lower1ex\hbox{$\sim$}}}}

\usepackage{xcolor}
\definecolor{red}{rgb}{1.0, 0, 0}

\begin{document}


\title{What the Tevatron Found?}
\author{Matthew R.~Buckley$^{1}$, Dan Hooper$^{1,2}$, Joachim Kopp$^3$, Adam Martin$^3$, and Ethan T.~Neil$^3$}
\affiliation{$^1$Center for Particle Astrophysics, Fermi National Accelerator Laboratory, Batavia, IL 60510}
\affiliation{$^2$Department of Astronomy and Astrophysics, University of Chicago, Chicago, IL 60637}
\affiliation{$^3$Theoretical Physics Department, Fermi National Accelerator Laboratory, Batavia, IL 60510}
\preprint{FERMILAB-PUB-11-359-A-T}
\date{\today}

\begin{abstract}
The CDF collaboration has reported a 4.1$\sigma$ excess in their lepton, missing energy, and dijets channel. This excess, which takes the form of an approximately Gaussian peak centered at a dijet invariant mass of 147 GeV, has provoked a great deal of experimental and theoretical interest. Although the D\O\ collaboration has reported that they do not observe a signal consistent with CDF, there is currently no widely accepted explanation for the discrepancy between these two experiments. A resolution of this issue is of great importance---not least because it may teach us lessons relevant for future searches at the LHC---and it will clearly require additional information. In this paper, we consider the ability of the Tevatron and LHC detectors to observe evidence associated with the CDF excess in a variety of channels. We also discuss the ability of selected kinematic distributions to distinguish between Standard Model explanations of the observed excess and various new physics scenarios.
\end{abstract}

\maketitle

\section{Introduction \label{sec:intro}}

Recently, the CDF collaboration reported the observation of an excess of events in their lepton, missing transverse energy (MET), and dijets channel~\cite{CDFCollaboration:2011fk,CDF73} (see also the thesis, Ref.~\cite{Cavalierethesis}). This excess appears among events with a dijet invariant mass in the range of 120-160 GeV, and is consistent with a Gaussian distribution centered at $147\pm 4$~GeV. As initially observed among 4.3~fb$^{-1}$ of data, the excess consisted of $156\pm42$ events in the electron channel and $97\pm38$ in the muon channel, constituting a 3.2$\sigma$ deviation from the Standard Model~\cite{CDFCollaboration:2011fk}. An excess was also identified in an inclusive analysis of events with a lepton, MET, and two or more jets, although with somewhat lower statistical significance. More recently, the CDF collaboration extended their analysis to 7.3~fb$^{-1}$ of data, finding excesses of $240\pm55$ and $158 \pm 45$ events in the electron and muon channels, respectively, and resulting in an overall significance of 4.1$\sigma$ (assuming a Gaussian shape for the signal)~\cite{CDF73}. 

If this excess is interpreted as evidence of a new particle decaying to a pair of jets, produced in association with a leptonically decaying $W^\pm$, then the observed rate requires a cross section on the order of $2$--$4$~pb, some 300 times larger than is predicted for processes including a Standard Model Higgs boson.\footnote{While the CDF and D\O\ collaborations have used a cross section of 4~pb as a canonical value for a new physics explanation, this is a very model-dependent quantity. Most models explaining the anomaly find that smaller ($\sim 2~$pb) cross sections can account for the observed events~\cite{Buckley:2011ly,Eichten:2011kx}. The discrepancy between these numbers seems to arise from larger detector acceptances for new physics models that have angular distributions and production mechanisms which are very different from the scaled-up Higgs boson used as a benchmark by CDF and D\O.} Unless this excess is the result of some not-yet-understood systematic effect (see Refs.~\cite{Campbell:2011uq,He:2011uq,Plehn:2011uq,Sullivan:2011fk} for possible explanations arising from errors in modeling of Standard Model physics\footnote{One of the initial proposals for a Standard Model explanation of the CDF excess was $t\bar{t}$ production~\cite{Plehn:2011uq}.  Semileptonic $t\bar{t}$ events can contribute to the dijets excess if two of the four jets are not identified, a process that is not well modeled perturbatively and is therefore sensitive to shower and detector effects. Also, although the dijet mass distribution for $t\bar{t}$ has a feature at $\sim$150 GeV, once showering is included, this feature is shifted to lower mass. Detector effects also move the feature toward lower masses, especially given the fact that flavor tagging is not used so $b$-jets are not specially treated~\cite{CDF73}. These mismodeling issues are not present in the inclusive analysis, which, as mentioned above, shows similar significance for the excess. For these reasons, we do not consider further this explanation for the CDF excess.}), it would constitute detection of physics beyond the Standard Model.

The initially enthusiastic response of the high energy physics community to CDF's excess has been tempered somewhat by the results of a subsequent study from the D\O\ collaboration~\cite{D0Collaboration:2011fk}. In their analysis of 4.3~fb$^{-1}$ of data, the D\O\ collaboration does not report a statistically significant excess consistent with that observed by CDF, but does favor (at the $\sim$1$\sigma$ level) the presence of a smaller bump-like feature at $\sim$150 GeV, corresponding to a production cross section of $0.82^{+0.83}_{-0.82}$ pb, or about 20\% as large as that reported by CDF. The D\O\ collaboration claims to exclude a `bump' arising from a 4~pb Higgs-like scalar with a statistical significance described by a $p$-value of $8\times 10^{-6}$~\cite{D0Collaboration:2011fk}. 
 
We note that the D\O\ analysis makes use of less than half of their current total data set, and does not contain any discussion of inclusive event distributions. The D\O\ analysis also differs from that of CDF in small, but potentially important ways. First, D\O\ uses a larger jet cone size ($R=0.5$ compared to $R=0.4$ used by CDF) which leads to a slower turn-on in $m_{jj}$ and a harder tail. Perhaps more importantly, the jet  energy calibration used by D\O\ includes a correction for `out-of-cone' radiation, thereby  attributing more energy to the reconstructed jet, while CDF does not include such corrections. As a result, a greater number of initial partons with less energy qualify as D\O\ jets compared to CDF jets. This is particularly  important given how sensitive the significance of CDF's excess is to the jet $p_T$ cut; relaxing CDF's cut from $p_T > 30$ GeV to $p_T > \sim 25$~GeV lowers the significance of the original 4.3 fb$^{-1}$ analysis from 3.2$\sigma$ to $\sim 2.5\sigma$, lowering the cut further to $p_T > 20$~GeV again reduces the significance to 1.1$\sigma$. It is not implausible that the difference in the way that D\O\ and CDF estimate jet energies could account for why D\O\ does not identify the large excess reported by CDF.

Given the discrepancy between the Tevatron's two experiments, more work is clearly called for. The differences between the results of CDF and D\O\ are highly unlikely to arise from statistical fluctuations, leaving only underlying systematic issues or actual new physics as possible resolutions. Even if the CDF excess is not a consequence of new physics, but is rather caused by some subtle mismodeling of the Standard Model backgrounds, it is possible that this error could propagate to the LHC experiments. It is therefore critical that the underlying cause(s) of the disagreement between the Tevatron experiments be definitively determined. The primary goal of this paper is to highlight additional channels and analyses at the Tevatron and the LHC which may be useful in this endeavor.

If the CDF anomaly is in fact caused by physics beyond the Standard Model, it is improbable that this physics is manifest in only the $\ell\nu+\text{dijet}$ channel. Of the twenty or so models that have been proposed to account for the CDF excess (see {\it e.g.} Refs.~\cite{Anchordoqui:2011kx,Babu:2011bh,Buckley:2011ly,Cao:2011qf,Chang:2011ve,Chen:2011fk,Cheung:2011uq,Dobrescu:2011fk,Dutta:2011cr,Eichten:2011kx,Enkhbat:2011fk,Fox:2011vn,Hewett:2011fk,Jung:2011lz,Jung:2011zr,Kilic:2011vn,Ko:2011kx,Liu:2011fk,Nelson:2011fk,Nielsen:2011uq,Sato:2011vn,Segre:2011fk,Wang:2011ys,Wang:2011yt,Yu:2011ve,Faraggi:2011uq,Ghosh:2011np,Anchordoqui:2011eg,Fan:2011vw,Harnik:2011mv}), all are predicted to produce associated signals in related channels, and most predict measurable deviations in the kinematic distributions of the reported CDF excess itself. Furthermore, while ATLAS and CMS do not have sufficient processed data to directly probe most models of new physics contributing to these channels, this will not be the case for long. There is currently more than 1.2~fb$^{-1}$ of LHC data recorded and being analyzed. As we will show, many of the new physics scenarios proposed to explain the CDF excess should become visible at the LHC with $\sim5$~fb$^{-1}$ of data, allowing for new cross-checks in the near future.

In this paper, we identify and discuss three different approaches to cross-check the CDF excess: comparisons of kinematic distributions of $\ell\nu+\text{dijet}$ events, searches in additional channels at the Tevatron, and searches at the LHC (see also Refs.~\cite{Cheung:2011vx,Eichten:2011fk,Harigaya:2011ww}). These tests serve not only to determine whether or not new physics is responsible for the CDF excess, but can also provide discriminating power between the various new physics scenarios which have been proposed. 

In Section~\ref{sec:models}, we describe in more detail the new physics models under consideration, focusing on $Z'$ and technicolor scenarios as representative examples. In Section~\ref{sec:kinematics} we compare the predictions of these models in the $\ell\nu+\text{dijet}$ channel to the CDF data, paying special attention to kinematic quantities other than the dijet invariant mass in which the excess was first discovered. We investigate the prospects of distinguishing different explanations of the excess, including a modified Standard Model background, on the basis of the kinematic distributions. In Section~\ref{sec:tevatron}, we calculate the cross sections and expected significance of new physics signals in alternative channels at the Tevatron experiments. As we will show, these channels will not only provide critical checks of the new physics interpretation of the CDF anomaly, but can also provide valuable information which can allow for discrimination between the various proposed models. In Section~\ref{sec:LHC}, we calculate the expected cross sections and estimate the luminosity required to observe an excess at the $3\sigma$ level (statistics-only) in each proposed channel (including $\ell\nu+\text{dijet}$) at the LHC. In Section~\ref{sec:conclusion}, we summarize the conclusions of our study and discuss the prospects for resolving the currently confusing situation.

\section{New Physics Models of the $W^{\pm}$ Plus Dijet Anomaly \label{sec:models}}

In this section, we briefly summarize some of the various models that have been proposed to explain the CDF dijet anomaly. Although these models differ significantly in terms of the underlying physics and high energy completion, we can broadly group the majority of them into two categories, each of which leads to the CDF excess through Feynman diagrams with distinct topologies. These categories are $t$-channel production (see, for example, Refs.~\cite{Buckley:2011ly,Enkhbat:2011fk,Nelson:2011fk,Segre:2011fk,Anchordoqui:2011kx,Babu:2011bh,Chang:2011ve,Cheung:2011uq,Fox:2011vn,Jung:2011lz,Jung:2011zr,Ko:2011kx,Liu:2011fk,Wang:2011ys,Yu:2011ve}), where a new particle of mass $\sim$140-160 GeV is produced in association with a $W^\pm$, and $s$-channel production (see, for example, Refs.~\cite{Eichten:2011kx,Cao:2011qf,Chen:2011fk,Kilic:2011vn,Ghosh:2011np,Fan:2011vw}), where a heavy particle is produced and decays into the lighter resonance seen in the dijet invariant mass distribution along with a $W^\pm$. 

Although these two categories of models cover the majority of new physics scenarios proposed to account for the CDF excess, there are others which do not fall into either of these categories~\cite{Dobrescu:2011fk,Wang:2011yt,Sato:2011vn}. These models often produce two new $\sim$$150$~GeV particles, one of which decays into a lepton plus a neutrino (mimicking a leptonically decaying $W^\pm$), and the other into a jet pair, which results in the invariant mass peak of the CDF anomaly. As such models can lead to very different phenomenology (and thus vastly different cross section predictions in other channels and at the LHC) depending on the detailed structure of the model, we must defer careful consideration to another time.

\subsection{$t$-Channel Models \label{sec:zprime}}

\begin{figure}[ht]
  \includegraphics[width=0.25\columnwidth]{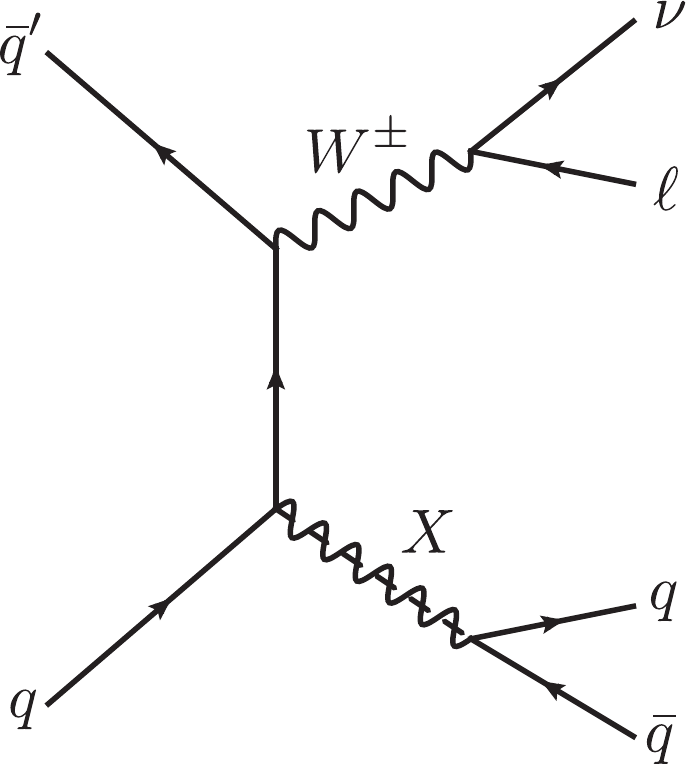}
  \caption{Representative Feynman diagram for the $t$-channel topology. $X$ in this case can be either a scalar or a vector boson. \label{fig:1resfeyn}} 
\end{figure}

Perhaps the simplest way to explain the CDF anomaly is to add a single new $\sim$140-160 GeV particle that couples predominantly to quarks. Such a state can be produced in association with a $W^\pm$ through diagrams similar to that shown in Fig.~\ref{fig:1resfeyn}. As quarks must be present in the initial states in order to produce a $W^\pm$, the new particle does not necessarily need to possess couplings to gluons.

The new particle, $X$, in this diagram could be either a scalar~\cite{Enkhbat:2011fk,Nelson:2011fk,Segre:2011fk} or a $Z'$ vector boson~\cite{Buckley:2011ly,Anchordoqui:2011kx,Babu:2011bh,Chang:2011ve,Cheung:2011uq,Fox:2011vn,Jung:2011lz,Jung:2011zr,Ko:2011kx,Liu:2011fk,Wang:2011ys,Yu:2011ve,Faraggi:2011uq,Anchordoqui:2011eg}. In either case, the couplings of this state to leptons must be significantly suppressed relative to those to quarks in order to evade constraints from LEP-II and the Tevatron~\cite{Buckley:2011ly,Yu:2011ve}. Throughout this study, we will assume negligible couplings to leptons. From the size of the observed $\ell\nu+\text{dijet}$ excess, the cross section for $W^\pm X$ production can be determined. From this, we can deduce information pertaining to the couplings of the $X$ to up and down quarks. For a representative choice of $t$-channel models, we will consider a $Z'$ scenario with an interaction Lagrangian given by:
\begin{equation}
\mathcal{L}_{Z'} \supset \sum_{q}  g_{Z'q}\, \bar{q}\, \gamma^{\mu}\, q \, Z'_{\mu},  
\end{equation}
where the sum is over both left- and right-handed quarks.

The couplings of a $Z'$ to various types of quarks and other fermions is somewhat model dependent. We will take a bottom-up approach (without assuming any particular UV-completion) which allows us the freedom to choose the couplings to the various left- and right-handed fields independently.\footnote{We assume anomalies are cancelled by new exotic fermions at somewhat higher scales. For explicit realizations of this, see Refs.~\cite{Buckley:2011ys,Dulaney:2010fk,Perez:2010uq}, in which baryon and lepton number are each gauged, or Ref.~\cite{Buckley:2011mm} which fits the $Z'$ into an $E_6$ grand unified group. Both connect the new $Z'$ to the dark matter of our universe.} Note that due to the chiral nature of quarks under $SU(2)_L$, only the left-handed quarks contribute to the CDF dijet excess. Although mechanisms exist which would allow the quarks within a left-handed doublet to have different effective couplings to the $Z'$~\cite{Fox:2011vn}, here we consider only cases in which the $Z'$ couples identically to the components of quark isodoublets. In particular, we consider two benchmark models: a ``left-handed'' model in which the couplings to the right-handed quarks are set to zero, while the couplings to the left-handed quarks are set at a canonical value chosen to explain the CDF excess, namely $g_{Z' q_L} = 0.3$~\cite{Buckley:2011ly,Cheung:2011uq,Jung:2011zr,Wang:2011ys,Yu:2011ve}, and a ``universal'' model, in which all quarks have the same coupling to the $Z'$, {\it i.e.}~$g_{Z' q_L} = g_{Z' u_R} = g_{Z'd_R}=0.3$. Although these two scenarios lead to an identical $\ell\nu+\text{dijet}$ signal, they yield different predictions for other channels. For example, a measurement of the cross section for the production of a $Z'$ in association with a photon will be approximately two times larger in the case of universal couplings than in the case of only left-handed couplings. As in Ref.~\cite{Buckley:2011ly}, we choose a $Z'$ mass of 150~GeV. This choice of parameters provides a good fit to the CDF data in the $W+\text{dijet}$ channel (see Fig.~\ref{fig:wjj} below).

We note that $t$-channel models which incorporate a scalar rather than a vector boson can more easily incorporate a non-trivial flavor structure \cite{Nelson:2011fk}.\footnote{It is also possible that a $Z'$ could have generation dependent couplings, especially given our poor understanding of flavor in the Standard Model. See, for example, Refs.~\cite{Buckley:2011ly,Fox:2011vn}.}  Although the additional freedom this would allow could result in a range of phenomenological characteristics that is not encompassed by our two $Z'$ benchmarks, it is impractical to cover the full range of all possible models here. Instead, we rely on our selected benchmark models to illustrate the kind of resolving power we expect future studies to have on the theoretical model space.

\subsection{$s$-Channel Models \label{sec:technicolor}}

Alternatively, the excess reported by CDF could originate from events in which the the $W^\pm$ and the $\sim$140-160 GeV $X$ state are produced through a heavier ($m_{X'} \gsim 250$ GeV) $s$-channel resonance. Realizations of this event topology can be found, for example, within the context of low-scale technicolor~\cite{Eichten:2011kx},\footnote{As technicolor models introduce additional fundamental fermions rather than bosons, due to Fermilab's location, Pauli's second exclusion principle gives some theoretical weight to this scenario over others~\cite{resonaances}.}
 R-parity violating supersymmetric models~\cite{Kilic:2011vn,Ghosh:2011np}, and quasi-inert two-Higgs doublet models~\cite{Cao:2011qf}.
In $R$-parity violating supersymmetric models, it is possible to produce sleptons and squarks on resonance through terms in the superpotential of the form $\lambda'_{ijk} L_i Q_j D_k^c$ or $\lambda''_{ijk} U_i^c D_j^c D_k^c$. Such a squark or slepton can then decay to a $W$ and a lighter squark or sneutrino, which then decays through the same R-parity violating coupling to the observed jets. In low-scale technicolor (LSTC) models~\cite{Lane:1989ej, Eichten:1996dx,Eichten:1997yq,Lane:1999uh,Mrenna:1999xj,Eichten:2007sx,Lane:2009ct}, the excess observed by CDF could originate from events in which a $\sim$300 GeV neutral (charged) technirho, $\rho_T$, is produced and then decays to a $W^\pm$ and a $\sim$150 GeV charged (neutral) technipion, $\pi_T$, which decays to jets (some fraction of which can be $b$-jets, depending on the CKM-like angles in the technicolor sector). 
 
\begin{figure}[ht]
  \includegraphics[width=0.32\columnwidth]{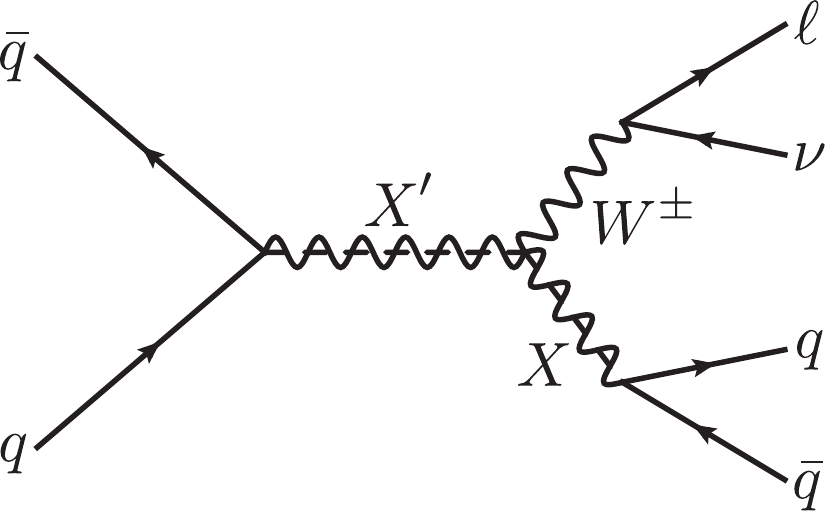}
  \caption{Representative Feynman diagram for the $s$-channel topology. $X$ and $X'$ in this case can be either scalars or vector bosons. \label{fig:2resfeyn}} 
\end{figure}

As the cross sections for the production of the dijet resonance in association with a $Z$ or $\gamma$ in this class of models is highly dependent on the charges of the new particles in question, the $Z+\text{dijet}$ and $\gamma+\text{dijet}$ channels will depend on the details of the underlying model. For explicit calculations, we will restrict ourselves to a single model: the LSTC model of Ref.~\cite{Eichten:2011kx}, with a 290~GeV $\rho_T$ decaying into a 160~GeV $\pi_T$ and a $W^\pm$. With these parameters, the model provides a good fit to the $W+\text{dijet}$ excess observed by CDF.

In LSTC, there are two different dynamically generated scales: $v_1$ and $v_2$, which both contribute to electroweak symmetry breaking. The sum of the scales in quadrature is fixed to the weak scale $v^2_1 + v^2_2 = v^2$, while the ratio $v_1/v_2$ is free. This allows a large separation in mass scales, and all the resonances which gain mass from the scale $v_1$ will be light. The fermion couplings to $\rho_T$ are small; effectively
\begin{equation}
g_{ff\rho_T} = g \Big(\frac{m_W}{m_{\rho_T}}\Big)\,\sin\chi,
\end{equation}
where $g$ is the Standard Model $SU(2)_L$ coupling and $\sin\chi$ is the fraction of the electroweak vacuum expectation value contained in the light condensate (analogous to $\tan\beta$ in two-Higgs doublet models). This factor can comfortably be $\sim$0.1, in which case resonances as light as 200-300~GeV are safe from all Tevatron and LEP bounds.

In addition to the $\rho_T$, the technicolor model also contains an isosinglet spin-1 partner $\omega_T$ as well as a spin-1, axial vector isotriplet $a_T$. We assume both have similar masses to the $\rho_T$, for reasons to be explained shortly. Neither the $\omega_T$ nor the $a_T$ contribute substantially to the $W+\pi_T$ decay channel. However they can contribute to other channels of interest; for example, those proceeding to $Z+\pi_T$ and $\gamma+\pi_T$ final states. In this study, we adopt the LSTC model of Ref.~\cite{Eichten:2011kx}, tuned to fit the $\ell\nu+jj$ excess. In particular, we set $m_{\omega_T} = m_{\rho_T}=290$~GeV, and $m_{a_T} = 1.1 \times m_{\rho_T}$.

For this choice of LSTC parameters, the $\rho_T$ is kinematically forbidden from decaying to a pair of technipions and must instead decay to at least one electroweak gauge boson. As the lighter sector only contributes a small amount to electroweak symmetry breaking, the couplings of the lightest $\rho_T$ are suppressed by factors of $v_1/v = \tan\chi$ ($\cong 1/3$ for the model used here) relative to the technirho-technipi-technipi coupling. Since the $\rho_T$ is forced to decay through suppressed couplings, it is very narrow, $\mathcal O(\text{GeV})$. The dominant decay mode is  $\rho_T \rightarrow \pi_T + W/Z$\,($\sim$70\%), followed by  $\rho_T \rightarrow  WW/WZ$\,($\sim$ 20\%). Decays to fermion pairs are possible, but are suppressed by additional powers of  $g_{ff\rho_T} $.

The choice of $m_{\omega_T} \sim m_{\rho_T}$ comes from the assumption that isospin violation is small and is driven by analogy with the near degeneracy of the QCD $\rho-\omega$ system~\cite{Lane:1989ej,Lane:2002wv}. The other mass choice, $m_{a_T} \sim m_{\rho_T}$, is quite unlike QCD. Instead, this relation is motivated by various arguments linking parity doubled spectra with a near conformal, or `walking' gauge coupling, a crucial ingredient in modern technicolor theories~\cite{Eichten:2007sx}. Parity doubling has also been argued to suppress contributions of new states to the electroweak parameter, $S$~\cite{Sundrum:1991rf, Lane:1994pg, Appelquist:1998xf,  Kurachi:2006mu, Appelquist:2010xv, Appelquist:2011dp}.

As neither the $\omega_T$ or $a_T$ can decay into (longitudinally polarized) electroweak gauge bosons, their decays are particularly susceptible to higher dimensional operators~\cite{Lane:1989ej, Eichten:2007sx}. The dominant operators are 
\begin{equation}
\frac{\kappa_1\, \epsilon_{ijk}\,\pi^i_{T}\, a^j_{T,\mu\nu}\,W^{k,\mu\nu}}{\Lambda_{TC}},\, \frac{\kappa_2\, a^{i,\lambda\mu}_T\,W_{i, \mu\nu}\,F^{\nu}_{ \lambda} }{\Lambda^2_{TC}},
\end{equation}
which lead to $a_T \rightarrow W + \pi_T$, $W + \gamma$, and
\begin{equation}
\frac{\kappa_3\, \pi^0_T\, \tilde{\omega}_{T,\mu\nu}\, F^{\mu\nu}}{\Lambda_{TC}},
\end{equation}
which leads to $\omega_T \rightarrow \pi^0_T + \gamma$. In these formulae, $i$ labels the isospin index and $\kappa_{1,2,3}$ denote the prefactors involving couplings and functions of $\chi$; see Ref.~\cite{Brooijmans:2008se} for a complete list of modes and prefactors. In these expressions, $a_{\mu\nu} (\omega_{\mu\nu})$ are the field strength tensors corresponding to the techni-$a$ (techniomega), $\tilde{\omega}$ is the dual field strength, and $W_{\mu\nu}$ and $F_{\mu\nu}$ are the $SU(2)_L$ and electromagnetic field strengths. Analogous higher dimensional operators can be written down for the $\rho_T$, but have less impact since they must compete with the $\pi_T+W/Z $ and $WW/WZ$ modes that proceed through marginal operators. The scale $\Lambda_{TC}$ is an additional parameter in the model and is assumed to be roughly the same scale as the masses of the resonances~\cite{Lane:1989ej,Eichten:1996dx,Eichten:1997yq,Lane:1999uh}. Finally, as with the $\rho^0_T$, the $a^0_T$ and $\omega^0_T$ can decay into Standard Model fermion pairs, although typically with quite small branching fractions, at least when $\Lambda_{TC} \sim m_{\rho_T}$~\cite{Lane:2009ct}.

The $Z'$ and technicolor models described above will serve as a basis for comparison between the $t$-channel and $s$-channel topologies in both the new channel searches and $\ell\nu+\text{dijet}$ kinematic distributions. Note that the slightly larger mass of the dijet resonance we have chosen for the technicolor model as compared to the $Z'$ model could result in observable effects once sufficient statistics are available. For the moment, however, both choices appear to be consistent with the CDF data. The overall cross section in the technicolor scenario is also somewhat larger than in the $Z'$ case, but again within the errors of CDF. We could, of course, decrease the cross section of the technicolor scenario to match the predictions of the $Z'$ models by increasing the $\rho_T$ and $\pi_T$ masses, but we choose to retain the parameters of Ref.~\cite{Eichten:2011kx}. Alternatively, we could increase the couplings of the $Z'$, but we refrain from doing so since the model would then come into some tension with existing constraints from dijet resonance searches at UA2~\cite{Buckley:2011ly,Yu:2011ve,Jung:2011zr,Cheung:2011uq}. For the parameters we have selected, both the $Z'$ and technicolor models provide good agreement with the data and are consistent with all existing constraints.

Lastly, we note that among $s$-channel models, the rates for associated production of the dijet resonance along with a $Z$ or photon are much more model dependent than in the $t$-channel case. In technicolor models, for example, the $Z$ and photon production processes are greatly enhanced by diagrams containing additional particles other than the $\rho_T$ that is required to produce the observed $\ell\nu+\text{dijet}$ signal. The Higgs model of Ref.~\cite{Cao:2011qf}, in contrast, contains no such additional states and thus does not predict a large rate of $Z$ associated production. $t$-channel models, in comparison, produce only one new particle through couplings to the Standard Model. As a result, predictions for signals appearing in other channels depend less on the specifics of the $t$-channel model under consideration.


\section{The $(W \to \ell\nu)+jj$ channel at the Tevatron \label{sec:kinematics}}

In this section, we investigate the models outlined in the Sec.~\ref{sec:models}, along with a more mundane explanation in terms of modified Standard Model backgrounds, in the context of the $(W \to \ell\nu)+jj$ channel. While the CDF excess in this channel was first observed at $3.2\sigma$ significance in a data set corresponding to only $4.3$~fb$^{-1}$ of integrated luminosity~\cite{CDFCollaboration:2011fk}, the analysis has been updated using $7.3$~fb$^{-1}$ of data, resulting in a increased significance of $4.1\sigma$~\cite{CDF73}. We remind the reader that the D\O\ collaboration's analysis of the $\ell\nu+\text{dijet}$ channel did not confirm the presence the excess reported by CDF~\cite{D0Collaboration:2011fk}. Here, we use the results of the $7.3$~fb$^{-1}$ analysis and compare the various kinematic distributions of the excess events (made available by the CDF collaboration in Ref.~\cite{CDF73}), to the predictions of the benchmark $Z'$ and technicolor models. Our goal here is to assess the prospects of distinguishing different explanations for the CDF anomaly directly in the $(W \to \ell\nu)+jj$ channel.

When simulating the $Z'$ models, we use a combination of Madgraph 5/MadEvent~\cite{Alwall:2011vn}, Pythia 6.4~\cite{Sjostrand:ys}, and Delphes~\cite{Ovyn:2009zr} for matrix element calculation, showering/hadronization, and detector simulation, respectively. The $Z'$ MadGraph models were constructed using Feynrules 1.6~\cite{Christensen:2008ly}. As technicolor has been previously implemented in Pythia 6, the model-building and MadGraph/MadEvent steps were unnecessary in this case, and we generated events directly in Pythia before passing them through Delphes. The CTEQ6.1L parton distribution function \cite{Stump:2003yu} was used for all simulations.

Since Delphes does not include an implementation of the CDF detector, we have created one, using information from publicly available sources \cite{Blair:1996kx,MovillaFernandez:2006vq,Hays:2005cj,triggertower}. We have tuned the resolution of the hadronic calorimeter in our simulation to optimally match the observed shape of the diboson peak in the dijet invariant mass distribution from the CDF $\ell\nu+\text{dijet}$ sample.

We apply the following cuts (which mimic as closely as possibly those used by CDF): A single lepton (electron or muon) with $E_T > 20$~GeV and $|\eta|<1.0$ is required, along with missing transverse energy MET$>25$~GeV. The transverse mass of the lepton+MET system (by assumption, a $W^\pm$) must be greater than $30$~GeV. Exactly two jets are allowed in the exclusive analysis, both of which must  have $E_T > 30$~GeV and $|\eta|<2.4$. CDF uses a fixed cone jet algorithm with a cone size of $\Delta R = 0.4$. There must be a separation of $\Delta R \geq 0.52$ between both jets and the lepton, $|\Delta \phi| >0.4$ between the MET and the leading jet, and $|\Delta \eta| <2.5$ between the two jets. The $p_T$ of the dijet system must be greater than $40$~GeV.

As a benchmark ``mis-modeling'' explanation of the CDF anomaly, we consider the possibility that the dominant Standard Model background (a $W^\pm$ produced in association with two jets from QCD interactions), is modeled incorrectly. We simulate this background using Alpgen~\cite{Mangano:2002ea}, Pythia and Delphes, and then reweight the resulting events such that their dijet invariant mass distribution forms a Gaussian bump centered around $m_{jj} = 140$~GeV and with a width of 20~GeV. This procedure emulates a hypothetical mis-modeling of the dominant Standard Model background. Even though careful investigations have not unearthed any evidence for such mis-modeling (or for any of the other proposed Standard Model explanations to date), we consider its inclusion as a useful exercise.

\begin{figure}
  \begin{center}
    \includegraphics[width=10cm]{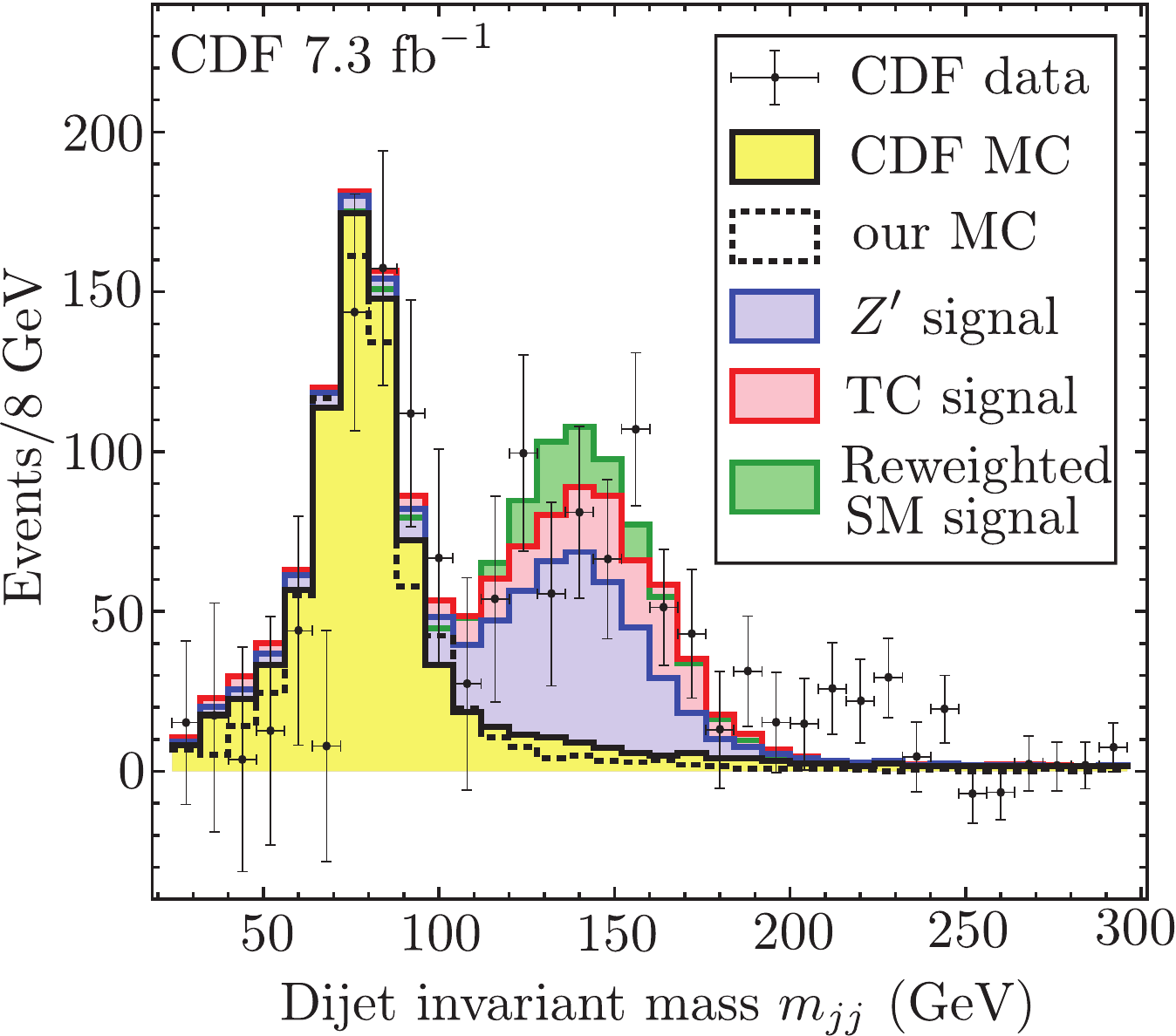}
  \end{center}
  \caption{Differential event distributions of the $Z'$ (blue) and technicolor (red) benchmark models plus the Standard Model electroweak diboson background (yellow) from~\cite{CDF73} as function of dijet invariant mass $m_{jj}$. The black data points with error bars show the CDF measurement using 7.3~fb$^{-1}$ of integrated luminosity~\cite{CDF73}, and the dotted black histogram is our own background prediction (which we use only to validate our simulations). The difference in normalization between the $Z'$ and technicolor peaks is due to our particular choice of parameters, which is explained in Sec.~\ref{sec:models}.}
  \label{fig:wjj}
\end{figure}

In Fig.~\ref{fig:wjj}, we show the distribution of the dijet invariant mass, $m_{jj}$, after subtraction of all Standard Model backgrounds except electroweak diboson production for our $Z'$ (blue) and technicolor (red) models, as well as for the reweighted Standard Model background (green). We compare these distributions to both the CDF data and the Standard Model prediction for 7.3~fb$^{-1}$ of integrated luminosity. Note the good agreement between our prediction for Standard Model diboson production and CDF's Monte Carlo expectation and measurement. We see from Fig.~\ref{fig:wjj} that each of these models is capable of explaining the observed excess. The $Z'$ prediction falls somewhat short of observed size of the excess, and the UA2 constraint~\cite{Buckley:2011ly} prevents us from choosing a significantly larger coupling constant. Given the uncertainties in the CDF measurement and the preference for small (or zero) production cross sections in D\O, however, we do not consider this observation alone to be a discriminant between these two models.

For each of the three new physics models considered ($Z'$ left-handed, $Z'$ universal, and technicolor), we summarize in the second column of Table~\ref{tab:tevatron} the cross sections before and after taking into account the $W \to \ell\nu$ branching ratio and the experimental cuts and efficiencies, and we also give an estimate for the signal-to-square root background ($S/\sqrt{N}$) of the excess expected in the full Tevatron Run II data set of 10~fb$^{-1}$. Here, the signal ($S$) and background ($N$) rates are those appearing over the region of $120\text{ GeV} < m_{jj} < 180\text{ GeV}$. We have verified that for both the $Z'$ and technicolor models, this choice of signal window is near-optimal.
The number of background events for 10~fb$^{-1}$ of data can be estimated to be $N=3815$ from the results of the CDF simulation shown in Ref.~\cite{CDF73}.

\begin{table}
  \parbox{10cm}{
  \renewcommand{\arraystretch}{1.2}
  \begin{ruledtabular}
  \begin{tabular}{lcccc}
                    & $\ell\nu+jj$ & $\ell\ell+j j$ & $\nu \nu +j j$ & $\gamma + j j$  \\ \hline

                    & 2400~fb      &    840~fb      &   840~fb       &     420~fb      \\
   $Z'$ left-handed &   41~fb      &    6.1~fb      &    23~fb       &      2.1~fb      \\
                    & $S/\sqrt{N}$\,=\,6.6  & $S/\sqrt{N}$\,=\,2.5    & $S/\sqrt{N}$\,=\,1.7    & $S/\sqrt{N}$\,=\,0.7    \\ \hline

                    & 2400~fb      &    970~fb      &   970~fb       &     840~fb      \\
   $Z'$ universal   &   40~fb      &     6.9~fb     &    25~fb       &      4.0~fb      \\
                    & $S/\sqrt{N}$\,=\,6.6  & $S/\sqrt{N}$\,=\,2.8    & $S/\sqrt{N}$\,=\,1.9    & $S/\sqrt{N}$\,=\,1.4     \\ \hline

                    & 2310~fb      &    530~fb      &   530~fb       &     541~fb      \\
   Technicolor      &   60~fb      &    6.8~fb      &    18~fb       &      2.8~fb      \\
                    & $S/\sqrt{N}$\,=\,9.7 & $S/\sqrt{N}$\,=\,2.8    & $S/\sqrt{N}$\,=\,1.4    & $S/\sqrt{N}$\,=\,1.1    \\
  \end{tabular}
  \end{ruledtabular}}

  \caption{Cross sections and expected signal-to-square root background ratios ($S/\sqrt{N}$) with 10~fb$^{-1}$ of Tevatron data for the three new physics models introduced in Sec.~\ref{sec:models}. The second column corresponds to the $WX \to \ell\nu+jj$ channel (with $X=Z'$ for the $Z'$ models and $X = \pi_T$ for technicolor), in which the excess was first observed and which was therefore used to choose the model parameters (see Sec.~\ref{sec:kinematics} for details). The third, fourth, and fifth columns correspond to the $ZX \to \ell \ell +j j $,  $ZX \to \bar\nu\nu +j j $, and $\gamma X \to \gamma + j j$ channels, respectively, which we will discuss in Sec.~\ref{sec:tevatron}. For each channel and model, we report three quantities: The cross section before branching ratios and experimental cuts, the cross section after application of branching ratios and cuts on final states, and the expected signal-to-square root background ratio in 10~fb$^{-1}$ of CDF data. For the $Z'$ models, even the pre-cut cross section in the $\gamma + j j$ channel includes a 60~GeV photon $p_T$ cut to avoid infrared divergences. See Sec.~\ref{sec:tevatron} for further details.}
\label{tab:tevatron}
\end{table}

Clearly, the signal-to-square root background ratio is only a very crude approximation to the statistical significance that an actual experimental analysis might obtain: on the one hand, it does not include systematic uncertainties. On the other hand, the CDF and D\O\ collaboration have demonstrated impressively in the past that more sophisticated analysis techniques---for instance multivariate approaches taking into account many kinematic variables at once---can greatly enhance their discovery potential. Therefore, we believe that by quoting signal-to-square root background ratios, we are not being unrealistically optimistic. Recall that the significance CDF reported from 7.3~fb$^{-1}$ of data (using a cut-based analysis and including systematic uncertainties) is $4.1\sigma$. For the same amount of data, we would obtain a significance of $S/\sqrt{N} = 5.4$ for the two $Z'$ models and 8.1 for technicolor.  The difference between the two models arises from our particular choice of parameters in the two models ($g_{Z'q_L}$ in the $Z'$ models, mass of the $\rho_T$ and $\pi_T$ in technicolor), the rationale for which has been given in Sec.~\ref{sec:models}. Although these numbers illustrate the limitations of the signal-to-square root background ratios quoted here, we believe that they are still useful for comparing different models and different search channels (see Sec.~\ref{sec:tevatron})\footnote{In this study, we also neglect theory-level systematic uncertainties coming from the scale dependence of NLO calculations of the background cross sections. As can be seen in Ref.~\cite{Campbell:2011gp}, these uncertainties are expected to be on the order of 10\%, and can thus be neglected here.}.

\begin{figure}
  \begin{center}
    \begin{tabular}{cc}
      \includegraphics[width=6cm]{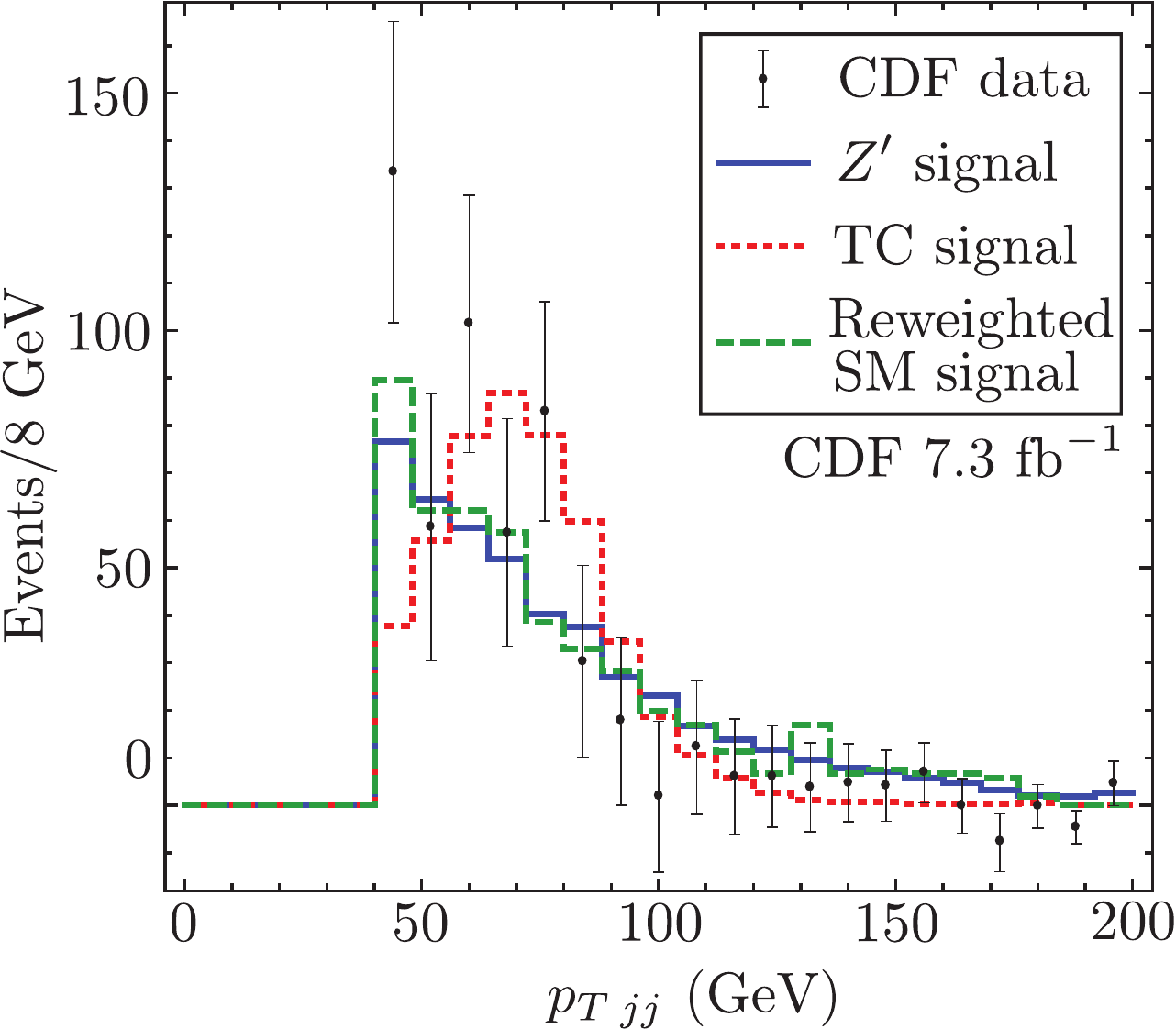} &
      \includegraphics[width=6cm]{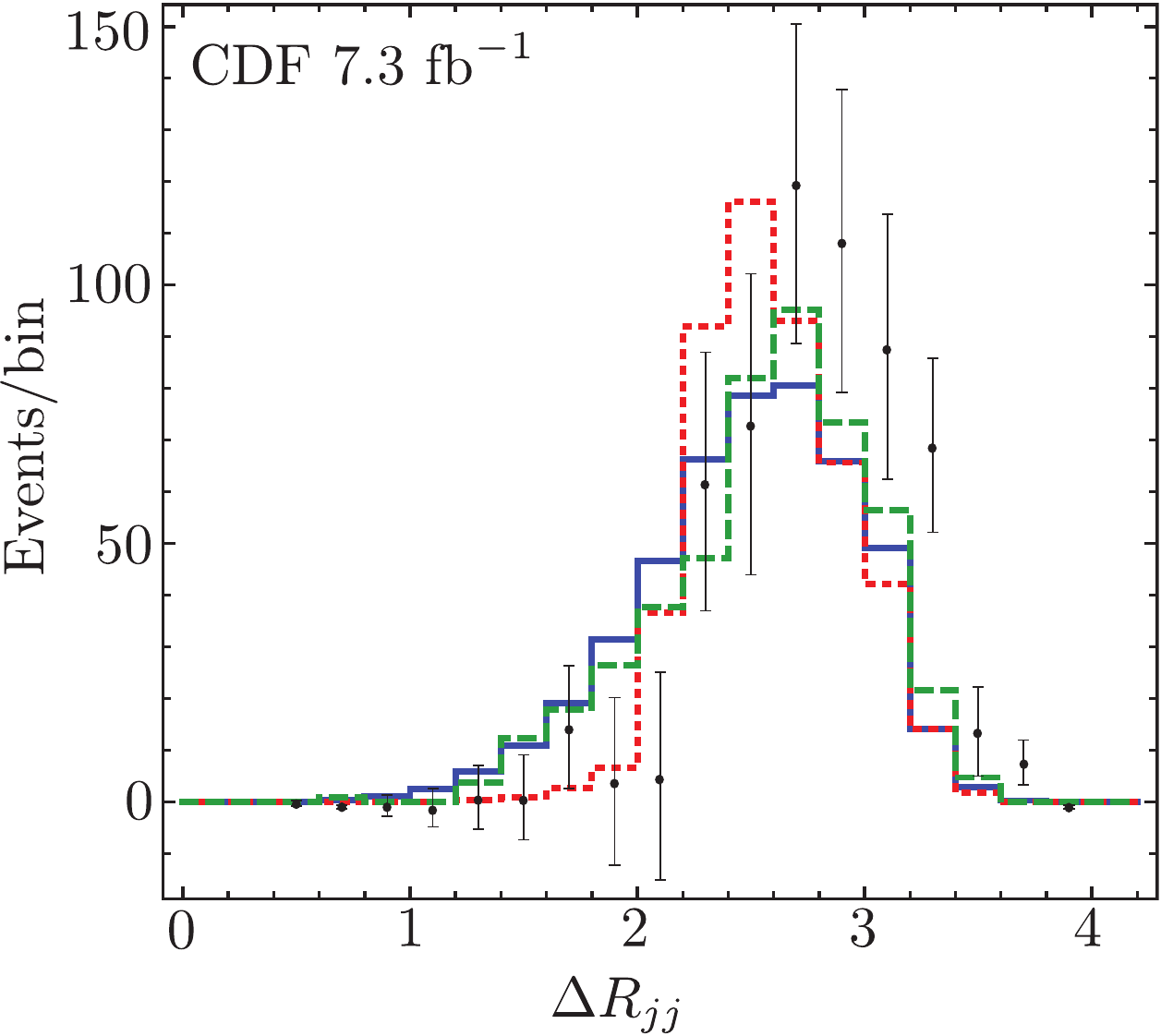} \\
      (a) & (b) \\
      \includegraphics[width=6cm]{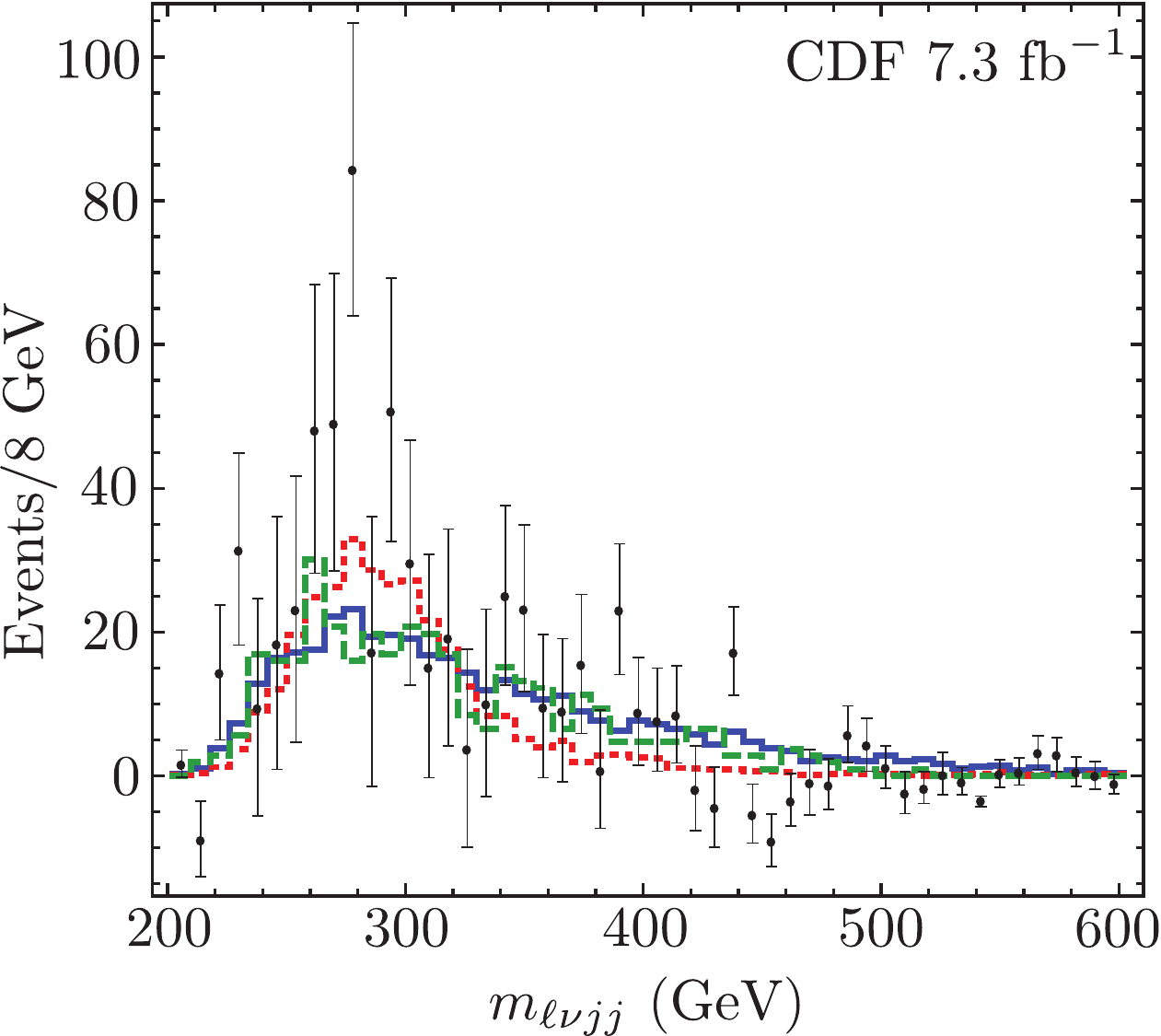} &
      \includegraphics[width=6cm]{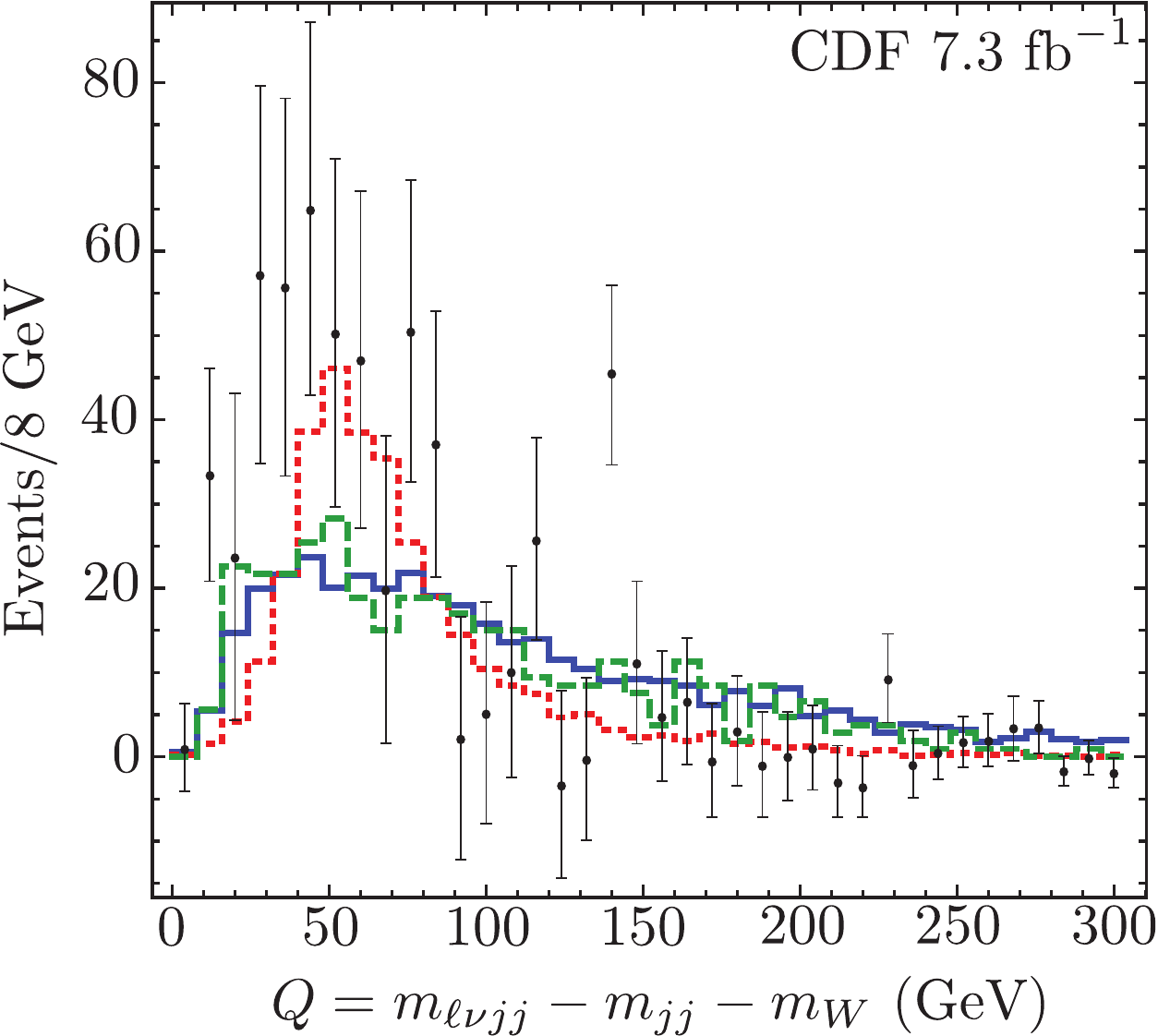} \\
      (c) & (d)
    \end{tabular}
  \end{center}
  \caption{Kinematic distributions for the $\ell\nu+\text{dijet}$ sample in 7.3~fb$^{-1}$ of CDF data after subtraction of Standard Model backgrounds (including diboson production). We show (a) the transverse momentum distribution of the dijet system, (b) the angular separation, $\Delta R_{jj}$, between the two jets, (c) the invariant mass of the $\ell\nu j j$ system, and (d) the variable $Q = m_{\ell\nu jj} - m_{jj} - m_W$. The black points with error bars denote the CDF data, while the solid blue, dotted red, and dashed green histograms correspond to the predictions of the $Z'$, technicolor, and reweighted Standard Model scenarios, respectively. In each case, we have normalized the model prediction to the number of observed excess events.}
  \label{fig:kinematics}
\end{figure}

Besides the $m_{jj}$ distribution shown in Fig.~\ref{fig:wjj}, we have studied a number of other kinematic distributions provided by the CDF collaboration~\cite{CDF73}. Among the sixteen kinematic variables presented by CDF, the ones that seem to have the greatest discriminating power between new physics models are the $p_T$ distribution of the jet pair ($p_{T,jj}$), the angular jet separation variable $\Delta R_{jj} = \sqrt{\Delta \eta_{jj}^2 + \Delta \phi_{jj}^2}$ (where $\Delta \eta_{jj}$ and $\Delta \phi_{jj}$ are the pseudo-rapidity and azimuthal separations, respectively), the invariant mass distribution of the $\ell jj + \text{MET}$ system ($m_{Wjj}$), and the quantity $Q = m_{Wjj} - m_{jj} - m_W$ (note that we assume the lepton and missing energy to come from an on-shell $W^\pm$). There is a two-fold ambiguity in the definition of $m_{Wjj}$ as the component of the neutrino momentum parallel to the beam axis, $p_{\nu,z}$, is not known and has to be calculated from a quadratic equation which can have two solutions. We follow CDF and always pick the smaller of the two $p_{\nu,z}$ values. In Fig.~\ref{fig:kinematics}, we plot each of these distribution and compare the results of CDF's 7.3~fb$^{-1}$ study to the predictions from the $Z'$ and technicolor models, as well as from the reweighted Standard Model background (recall that there is no difference between the left-handed and universal $Z'$ models in the $\ell\nu + \text{dijets}$ channel). Since the normalization in all models can be adjusted to a certain degree, we focus here on the shape of these kinematic distributions as a more important discriminant. Consequently, we have normalized all model predictions in Fig.~\ref{fig:kinematics} to the number of observed excess events.

Indeed, we find several intriguing shape differences between the different models: The $p_{T,jj}$ distribution, as shown in Fig.~\ref{fig:kinematics}(a), is much flatter for the Standard Model and $Z'$ cases, and it has a sharp turn-on at the $p_{T,jj}$ threshold of 40~GeV for these models, whereas the same distribution for technicolor exhibits a peak. The reason for this is that, in technicolor, the $p_{T,jj}$ of relatively central jets cannot be much smaller than $m_{\rho_T} - m_{\pi_T} - m_W$. Even though such a peak is not currently visible in the data, any resulting tension is very mild because of the large statistical uncertainties.

On the other hand, the shapes of the technicolor predictions for the $\Delta R_{jj}$, $m_{Wjj}$, and $Q$ distribution match the data quite well (see Fig.~\ref{fig:kinematics}(b), (c), and (d)), except for a slight shift in the $\Delta R_{jj}$ peak towards lower values. In the technicolor model (or any other $s$-channel model), one expects a peak to appear at the mass of the heavy resonance in the total invariant mass, $m_{Wjj}$, of the lepton, missing energy and dijet system. This feature is less prominent than the peak in the $m_{jj}$ distribution due to both the ambiguity in $p_{\nu,z}$ and the large Standard Model background in the relevant region. In the $Q$ distribution, technicolor predicts a peak located around the difference between $m_{\rho_T}$ and the sum of $m_{\pi_T}$ and $m_W$, which, for our choice of parameters, corresponds to $\sim 50$~GeV. This peak should be sharper than the peaks in the $m_{Wjj}$ or $m_{jj}$ distributions alone, since the jet energy uncertainty, which dominates the widths of these distributions, cancels in the difference. No sharp features are expected in any of the distributions discussed here for a $t$-channel model or for the modified Standard Model background.

These arguments help us to understand the results shown in Fig.~\ref{fig:kinematics}, although some of the predicted features (for instance the technicolor peak in $m_{Wjj}$) are hidden beneath the background. From these plots, we conclude that although kinematic distributions may provide an interesting discriminant between models, no firm conclusions can be drawn with the present statistics. It is possible, however, that a multivariate analysis, combining all kinematic distributions and using the full 10~fb$^{-1}$ of data that CDF and D\O\ will have at the end of Run II, might be able to significantly favor or disfavor some theoretical explanations of the dijet excess.

\section{Alternative Search Channels at the Tevatron\label{sec:tevatron}}

Whatever the true explanation of the CDF excess is, many theoretical models predict that it will also manifest itself in one or more of the related channels: $(Z \to \ell^+\ell^-)+jj$, $(Z \to \bar\nu\nu)+jj$, and $\gamma+jj$. In this section, we will study each of these channels and make predictions for their cross sections and observability in the $Z'$ and technicolor models.

\subsection{$(Z \to \ell^+\ell^-)+jj$}

We begin by considering the $\ell^+\ell^-$+dijet search channel, where the two leptons are assumed to be the decay products of a Standard Model $Z$. While a dedicated CDF analysis of this channel is available \cite{leptonleptondijet}, it does not provide a background distribution for the dijet invariant mass. We therefore use the $(Z \to \ell^+ \ell^-)+\text{dijet}$ analysis from Ref.~\cite{Cavalierethesis}, which is based on 4.3~fb$^{-1}$ of data. The cuts for the jets and highest $p_T$ lepton are as in the $\ell\nu+\text{dijet}$ analysis described previously; the second lepton must be opposite sign, same flavor as the first, and have $p_T >10$~GeV. The dilepton invariant mass must be in the range $81~\mbox{GeV} < m_{\ell \ell}<110~\mbox{GeV}$. 

As we did for $(W \to \ell\nu)+jj$, we select a $60$~GeV-wide window in the dijet invariant mass distribution.  In the reported background for 4.3~fb$^{-1}$, CDF expects 255 events in this window (combined electron and muon channels), and so we estimate a background of 593 events for a search in 10~fb$^{-1}$. In Table~\ref{tab:tevatron}, we report the cross section (both before and after application of cuts and the $Z \to \ell^+\ell^-$ branching ratio) and the signal-to-square root background ratio ($S/\sqrt{N}$) for each benchmark model.

For all three of the models considered, we see that this channel holds a great deal of promise. If a full analysis of the currently available Tevatron data set sees no dijet resonance in this channel, then we would be able to say, with $\sim$$90\%$ confidence, that models akin to the $Z'$ or technicolor scenarios cannot be responsible for the CDF excess. As these models are not particularly unique in their predictions of sizable $Z+X$ production cross sections, the lack of a dijet resonance in leptons+dijets would constitute strong evidence against a new physics explanation of the CDF anomaly. It would, however, leave room for $s$-channel models with no associated production of $Z$ bosons (see, for example, Ref.~\cite{Cao:2011qf}). Additionally, as a dijet bump arising from the mismodelling of backgrounds or detector effects in the distribution of $l\nu+jj$ events would also be expected to lead to a dijet bump in the $ll+jj$ channel, the absence of such a feature would disfavor many Standard Model explanations for the CDF excess as well.

If new physics is responsible for the CDF excess, our analysis indicates that both the CDF and D\O\ data sets might be necessary for a $>3\sigma$ observation. However, as discussed previously, the $S/\sqrt{N}$ values given in Table~\ref{tab:tevatron} are based solely on the total number of signal and background events in the signal window, and the significance attainable in a full analysis could be somewhat worse due to systematic uncertainties, or somewhat better with more sophisticated analysis techniques (for instance multivariate methods taking into account the full kinematics of each event). In addition, while we see that the universal $Z'$ and technicolor models do have larger production cross sections than the left-handed $Z'$, the difference appears to be too small to allow for a discrimination between these models in this channel at the Tevatron.

\subsection{$(Z \to \bar\nu\nu)+jj$}

Next, we replace the $Z$ decaying to charged leptons with a $Z$ decaying to neutrinos, and consider the MET+dijet final state. Our analysis of this channel is based on a CDF search using $3.5$~fb$^{-1}$ of data~\cite{Aaltonen:2009fd,CDFmET}\footnote{A second CDF search in this channel can be found in Refs.~\cite{CDFmET2,Collaboration:2009fk}, while a similar D\O\ search (requiring $b$-tagged jets, however) is described in Refs.~\cite{D0mETb,D0mETb2}}. Following CDF, we require exactly two jets with $E_T > 25$~GeV, $|\eta|<2.0$, and with less than 90\% of the total energy of each jet deposited in the electromagnetic calorimeter. Events with more than two jets fulfilling these requirements are rejected.  The missing transverse energy has to be larger than 60~GeV, and has to be separated by $|\Delta\phi| > 0.4$ from either jet.  CDF imposes in addition a cut on the ``MET significance" and a cut on the fraction of total event energy deposited in the electromagnetic calorimeter, which helps to reduce non-collision backgrounds from cosmic rays, beam halo interactions, etc.  This last cut could not be accurately implemented in our detector simulation, and so was omitted in our signal calculations.

As with the dilepton+dijets channel, we estimate the signal-to-square root background ratio, $S/\sqrt{N}$, in a window in dijet invariant mass centered on $150$~GeV. As the reported backgrounds only extend to $160$~GeV, we must truncate our window slightly: from $120$~GeV up to 160~GeV (rather than 180~GeV). The available analysis of 3.5~fb$^{-1}$ predicts 6076 events in this window, corresponding to 17,361 background events in 10~fb$^{-1}$. For each benchmark model, we report the signal cross section (both before and after $Z\to \nu\bar{\nu}$ branching ratio and cuts) and the expected signal-to-square root background ratio in Table~\ref{tab:tevatron}. 

Although the signal cross sections are an order of magnitude higher in the missing energy channel than in the dilepton channel, the background is also significantly larger. Therefore, we do not find $2\sigma$ deviations from the Standard Model for any of our benchmark models. However, with improved analysis techniques, or a combined CDF--D\O\ data set, this channel may plausibly provide a useful cross-check. As with dilepton+dijets, we again do not see significant observable differences between our benchmark models.

\subsection{$\gamma+jj$}

Finally, we consider events with a high $p_T$ photon produced in association with two jets. This channel has been much discussed in the literature in light of the CDF anomaly. Unfortunately, many of the conclusions have been misleading (see, for example, the early versions of Refs.~\cite{Buckley:2011ly,Jung:2011zr}, the former by four of this paper's authors). In particular, in the online version of CDF's search in this channel~\cite{CDFgammajj}, the normalization of the Standard Model backgrounds were mislabeled by a factor of twenty~\cite{gammapersonal}. Additionally, the simulated background currently reported does not include jets from a $W^\pm$ or $Z$. These Standard Model backgrounds are significantly larger than the event rate predicted by any new physics models. At the moment, no significant bounds on the models considered here can be derived from the available studies in this channel.

Despite these difficulties, we can still ask what signals might be found in the Tevatron's full 10~fb$^{-1}$ data set. As the cuts previously used by CDF result in an extremely large Standard Model background, we strengthen the photon cut, requiring $E_T> 60$~GeV (instead of $E_T>30$~GeV) and $|\eta|<1.1$, as well as at least two jets each with $E_T > 30$~GeV and $|\eta|<2.4$.  We also add a photon isolation cut, rejecting any event with $\Delta R < 0.6$ between the photon and either of the two leading jets.  Finally, we mimic the previous searches and include additional cuts on the dijet system, requiring $|p_{T,jj}| > 40$ GeV and $|\Delta \eta| > 2.5$ between the jets.  If missing energy appears in the event within $\Delta \phi < 0.4$ of any jet, the event is rejected.

As the reported CDF background is problematic, we rely on Alpgen \cite{Mangano:2002ea} to simulate Standard Model background in this channel (again using CTEQ6.1L). Here and throughout the paper we use the default Alpgen scale choice of $Q^2 = p_{T,V}^2+\sum p_{T,j}^2$, where $p_{T,V}$ is the $p_T$ of the vector boson. The matching scale is $p_{T,\rm min}+5~$GeV. As before, we select a 60~GeV window centered on the reported peak in the invariant mass of $\sim$$150$~GeV. Using these cuts, we find a Standard Model background cross section of $78$~fb in the signal region, corresponding to 775 events in the 10~fb$^{-1}$ data set. We simulate the signal for each model as described previously, and report the resulting cross sections and signal-to-square root background ratios in the last column of Table~\ref{tab:tevatron}. As the $Z'+\gamma$ cross section has an infrared divergence when the photon $p_T$ goes to zero, even the `pre-cut' cross sections reported for the two $Z'$ models include already a $p_T > 60$~GeV cut on the photon. No such cut is placed on the pre-cut cross section for the technicolor model, and so some care must be taken in directly comparing the pre-cut cross sections for the different models in the $\gamma + j j$ channel.

Due to the immense background rate, no statistically significant deviations from the Standard Model are predicted. This is especially unfortunate as the photon channel provides, in principle, the best discrimination between the models under consideration, as can be seen from the vastly different $Z'$ cross sections in Table~\ref{tab:tevatron}. It should be noted that, barring the $\ell\nu+\text{dijet}$ channel, none of these Tevatron analyses have been optimized to look for a new particle at $\sim$$150$~GeV over the Standard Model background. It is possible that these prospects could be improved by the application of stricter cuts (see Ref.~\cite{Eichten:2011fk}).

\section{LHC Search Channels \label{sec:LHC}}

In this section, we will consider the ability of the ATLAS and CMS experiments at the LHC to probe the $\ell\nu+\text{dijet}$ channel in which the CDF anomaly was initially reported, as well as the additional channels considered in the previous section. We assume an LHC center of mass energy $\sqrt{s} = 7$~TeV. As before, we use the three benchmark models outlined in Sec.~\ref{sec:models}, and calculate the expected cross section (before and after branching ratios and experimental cuts are applied) for each. Unlike Sec.~\ref{sec:tevatron}, we do not calculate the signal-to-square root background ratio after a set luminosity, but rather determine the required luminosity for each experiment to observe a $S/\sqrt{N} = 3$ deviation from the Standard Model. In each case, we set an acceptance window of 60~GeV around the nominal center (150~GeV) of the mass peak in the dijet invariant mass and calculate $S/\sqrt{N}$. As before, we must neglect systematic effects and consider only statistical errors. As published searches are not available for most channels, we must extrapolate acceptable cuts and simulate backgrounds using publicly available tools. 

The $\ell\nu+\text{dijet}$ channel at the LHC is clearly the first place to look for evidence of the CDF anomaly. Studies for $(W^\pm \to \ell \nu)+(\text{1--4})$ jets are available from ATLAS (using 1.3~pb$^{-1}$ of data) \cite{ATLASWdijet,Collaboration:2010kx}, and the $(W^\pm \to \ell \nu) + n\text{ jets}$, $(Z\to\ell^+\ell^-)+n\text{ jets}$ final states have been studied at CMS (with 36~pb$^{-1}$ of data) \cite{CMSWdijet}. A more recent study from ATLAS, presented at EPS \cite{EPSATLAS}, used 1.02 fb$^{-1}$ of data and cuts modeled on CDF. As we will show, our benchmark models are not expected to be visible in this data set, though as Ref.~\cite{Harigaya:2011ww} demonstrates, models with predominant coupling to 2$^{\rm nd}$ and 3$^{\rm rd}$ generation quarks or gluons can be constrained. For this study, we will use the CMS lepton selection criteria, requiring exactly one charged lepton to pass the cuts.  For electrons, the requirements are $E_T > 20$~GeV and $|\eta| < 2.5$, but not in the range $1.4442-1.566$.  Muons must have $p_T > 20$~GeV and $|\eta|<2.1$.  The MET must be greater than $25$~GeV, and the event must reconstruct the lepton plus MET system with a transverse mass $m_T > 30$~GeV.  Exactly two jets are required, with $|p_T| > 30$~GeV and $|\eta| < 2.4$.  Jets are discarded if found within $\Delta R < 0.52$ of the identified lepton.  Following the CDF analysis in the same channel, we require $|p_{T,jj}| > 40$~GeV, and $\Delta \eta < 2.5$ between the two jets.  Finally, the event is discarded if $\Delta \phi < 0.4$ between the missing energy and the leading jet.

W simulate backgrounds using Alpgen, Pythia, and Delphes (using the CMS detector implementation and CTEQ6.1L). We require exactly two jets to pass the previously defined cuts (rather than 1--4). In the signal window $120~\mbox{GeV} < m_{jj} < 180$~GeV, we find a background cross section of $15$~pb. For each of our benchmark models, we calculate the signal cross section using the simulation chain of MadGraph/Pythia+Delphes described in Sec.~\ref{sec:tevatron}. We then calculate the required luminosity which will result in $S/\sqrt{N} > 3$. The results are found in Table~\ref{tab:LHC}. We find that it should be possible to rediscover or exclude the CDF anomaly before the first long shutdown of the LHC in 2013. Even the 2011 data set could plausibly yield a detection in this channel at the $3\sigma$ level. We emphasize again, however, that we cannot reliably estimate systematic errors, and have used a rather simplistic statistical approach. We have high hopes that the ATLAS and CMS experimental community will greatly improve on these techniques. 

Turning to the $\ell\ell+\text{dijet}$ channel, we apply the selection cuts from the CMS study \cite{CMSWdijet} for leptonically decaying $W/Z$, requiring exactly two opposite sign, same flavor leptons and exactly two jets. If the leptons are electrons, there must be one that passes the `tight' cuts of Ref.~ \cite{CMSWdijet}: $E_T > 20$~GeV, $|\eta|<1.4442$ or $1.56<|\eta|<2.500$, along with additional isolation cuts that we take to be 80\% efficient. The muon cuts require one muon to have $p_T > 20$~GeV and $|\eta|<2.1$. We do not apply the CMS isolation cut, as this has a negligible effect on the relavant backgrounds (see Ref.~\cite{CMSWdijet2} for details). The second lepton must have $p_T>10$~GeV, and (for electrons) must also pass a looser set of isolation cuts which we take to be 95\% efficient. The lepton pair must have an invariant mass between $60~\mbox{GeV} < m_{\ell \ell}<120$~GeV. The jets must have $p_T> 30$~GeV, $|\eta|<2.4$, and are discarded if within $\Delta R < 0.52$ of a tagged lepton.  

With these cuts, we again simulate the background using Alpgen, requiring the invariant mass of the jet pair to fall within our 60~GeV-wide signal window. We find a background cross section of $1.6$~pb. Signal events are then generated for each model, and the resulting cross sections are reported in Table~\ref{tab:LHC}. As before, we then calculate the required luminosity to have $S/\sqrt{N} > 3$.

We are not aware of an applicable published search strategy from either ATLAS or CMS that can be easily applied to either the photon + dijets or missing energy + dijet channels. We must therefore extrapolate from the previously discussed searches. For both $\gamma+\text{dijet}$ and MET+dijets we will require two jets with $p_T>30$~GeV and $|\eta|<2.4$, and with less than 90\% of their energy deposited in the electromagnetic calorimeter. In the missing energy channel, we require MET $> 60$~GeV, and also require a separation of $\Delta \phi > 0.4$ between the missing energy and both jets. For the photon channel, we require $p_T > 60$~GeV, and apply other cuts on the photon and jets exactly as described for the Tevatron photon-channel analysis. Again, using Alpgen, we calculate that the signal window has a background of $539$~fb for $\gamma+\text{dijet}$, and 7.1~pb for MET+dijets. It should be noted that, in the MET channel, our background simulation is likely underestimating the true rate. Firstly, our simulation cannot reliably model QCD multi-jet contributions to the background, which could be significant. Secondly, while cross-checking the validity of our $(Z \to \nu \nu) +jj$ simulation against the results in Ref.~\cite{daCosta:2011qk} (which is sensitive to a very different kinematic regime than the one selected in our analysis), we found our Alpgen Monte Carlo backgrounds which to be approximately $60\%$ of the ATLAS simulation. These caveats should be kept in mind when considering our results for the MET channel. However, even if these effects are quite large, this channel appears to be very promising. For all channels, signal cross sections and required luminosities are reported in Table~\ref{tab:LHC}.

\begin{table}[ht]
  \parbox{10cm}{
  \renewcommand{\arraystretch}{1.2}
  \begin{ruledtabular}
  \begin{tabular}{lrrr@{\hspace{-0.6cm}}p{0.1cm}@{\hspace{-0.2cm}}r}
                    & $\ell\nu+jj$ & $\ell\ell+j j$ & $\nu \nu +j j$ && $\gamma + j j$  \\ \hline

                    & 11400~fb     & 3400~fb        & 3400~fb        && 3450~fb         \\
   $Z'$ left-handed &    145~fb     &  13.7~fb        &   99~fb        &&   5.3~fb         \\
                    & 6.4~fb$^{-1}$& 75~fb$^{-1}$  & 6.5~fb$^{-1}$ & $^*$  & 170~fb$^{-1}$   \\ \hline

                    & 11400~fb     & 3800~fb        & 3800~fb        && 6900~fb         \\
   $Z'$ universal   &    143~fb     &  14.4~fb        &   106~fb        &&   11.9~fb         \\
                    & 6.6~fb$^{-1}$& 67~fb$^{-1}$  & 5.7~fb$^{-1}$  & $^*$  & 34.4~fb$^{-1}$   \\ \hline

                    & 7970~fb      & 2200~fb        & 2200~fb        && 1870~fb         \\
   Technicolor      &  188~fb      &  18.8~fb        &   75~fb        &&   6.9~fb         \\
                    & 3.8~fb$^{-1}$& 40~fb$^{-1}$  & $11.3$~fb$^{-1}$ & $^*$ & $103$~fb$^{-1}$ \\
  \end{tabular}
  \end{ruledtabular}}

  \caption{Cross sections (before and after branching ratios and experimental cuts) and required luminosity for signal-to-square root background ratio of 3 at the 7~TeV LHC for the three new physics models introduced  in Sec.~\ref{sec:models}. The search channels we consider are the same as in Table~\ref{tab:tevatron}. For the $Z'$ models, the pre-cut cross section in the $\gamma + j j$ channel again includes a 60~GeV photon $p_T$ cut to avoid infrared divergences. Our simulation of the $\nu\nu+jj$ channel may underestimate the background, this is indicated by an $*$ in the Table. See text of Sec.~\ref{sec:LHC} for further details.}
  \label{tab:LHC}
\end{table}

\section{Discussion and Conclusions \label{sec:conclusion}}

The observation of a high-significance excess in the $W + jj$ channel by the CDF collaboration, and the subsequent non-confirmation of this result by D\O\, have left the collider physics community in a very unsatisfactory situation. On the one hand, scientific conservatism dictates that a spectacular result should be confirmed independently before it is to be taken to be something approaching scientific fact. On the other hand, the CDF analysis has been extensively reviewed by a large number of experts, both within and outside of the collaboration, and no clear mistakes or problems have been identified. We believe that a resolution of the disagreement between CDF and D\O\ is of utmost importance for the high energy physics community. At present, we hold the position that an experimental result such as the CDF excess should not be discarded prematurely, just as it should not be accepted prematurely. Furthermore, whether the anomaly is due to new physics or has a more mundane explanation within the Standard Model, resolving it will provide important insights that will help prevent similar problems in future analyses.

In this paper, we have outlined several possible paths for clarifying this situation. We have first considered the $W+jj$ channel itself, and have stressed the discriminating power of kinematic variables other than the dijet invariant mass. As benchmarks, we have considered an ad-hoc modified Standard Model background, a model in which the dijet bump arises from the decay of a $Z'$ boson, and a low-scale technicolor scenario. We have identified interesting differences in the kinematic distributions predicted in these three cases and, even though low statistics prevents us from drawing any firm conclusions at this stage, we expect that a multivariate analysis using the full Tevatron dataset may be able to discriminate between these scenarios.

If the CDF anomaly is indeed the result of new physics, it will be very interesting to study final states different from, but related to, $W+jj$. At the Tevatron, we find that the $(Z \to \ell^+\ell^-) + jj$ channel should provide a signal observable at the $\sim$$3\sigma$ level if either a $Z'$ or technicolor model is responsible for the observed $W+jj$ excess. At the LHC, an excess in $l\nu+jj$ should become statistically significant by the end of the year if new physics is responsible for CDF's anomaly. By the end of 2012, a signal could also become observable in the $(Z \to \bar\nu\nu) + jj$ and $(Z \to \ell^+\ell^-) + jj$ channels.

\section*{Acknowledgments}

In the course of this work, we have benefited from discussions with many of our colleagues. We would like to thank the CDF collaboration, in particular Viviana Cavaliere, Ray Culbertson, and Sam Hewamanage for information on the $W+jj$ and $\gamma+jj$ analyses. Moreover, we are indebted to Johan Alwall and the MadGraph team, Bogdan Dobrescu, and Ciaran Williams for many useful discussions on theoretical issues related to the CDF excess. Fermilab is operated by Fermi Research Alliance, LLC, under Contract DE-AC02-07CH11359 with the United States Department of Energy.

\bibliography{Zprimepapers5}
\bibliographystyle{apsrev}
\end{document}